\documentclass[acmsmall,screen]{acmart}
\usepackage{multirow}
\usepackage{graphicx}
\usepackage{subcaption}
\usepackage{xspace}
\usepackage{algorithm}
\usepackage[noend]{algpseudocode}
\usepackage[normalem]{ulem}
\useunder{\uline}{\ul}{}
\usepackage{ragged2e}
\usepackage{array}
\usepackage{amsmath}
\usepackage{xcolor}
\usepackage{balance}
\usepackage{booktabs}
\usepackage[flushleft]{threeparttable}
\usepackage{threeparttable}
\usepackage{enumitem}
\usepackage{pifont}

\algrenewcommand\textproc{\texttt}

\usepackage[compact]{titlesec}
\titlespacing{\section}{5pt}{6pt}{5pt}
\AtBeginDocument{                    
  \setlength\abovedisplayskip{0pt}
  \setlength\belowdisplayskip{0pt}}
\algrenewcommand\textproc{\texttt}

\newcommand{\Answer}[2]{\noindent \textbf{Answer to #1:} #2}

\newcommand{\toolname}[0]{{\textsc{SAIL}}}

\newcommand{\appNum}[0]{35}
\newcommand{\testNum}[0]{104}

\newcommand\toller[1]{\textsc{Toller}}
\renewcommand{\Comment}[1]{}

\newcolumntype{C}[1]{>{\centering\arraybackslash}p{#1}}

\definecolor{customblue}{RGB}{192, 208, 235}
\definecolor{customdarkblue}{RGB}{68, 114, 196}

\usepackage{todonotes} 

\newif\ifshowtodos
\showtodostrue  

\AtBeginDocument{
  \providecommand\BibTeX{{
    \normalfont B\kern-0.5em{\scshape i\kern-0.25em b}\kern-0.8em\TeX}}}

\begin{document}

\title[]{Skill-Adaptive Imitation Learning for UI Test Reuse
}

\author{Mengzhou Wu}
\affiliation{
  \institution{Peking University}
  \city{Beijing}
  \country{China}
}
\email{wmz@stu.pku.edu.cn}

\author{Hao Wang}
\affiliation{
  \institution{UC Berkeley}
  \city{Berkeley}
  \country{USA}
}
\email{hwang628@berkeley.edu}

\author{Jun Ren}
\affiliation{
  \institution{University of Texas at Dallas}
  \city{Richardson}
  \country{USA}
}
\email{jxr210020@utdallas.edu}

\author{Yuan Cao}
\affiliation{
  \institution{Peking University}
  \city{Beijing}
  \country{China}
}
\email{cao_yuan21@stu.pku.edu.cn}

\author{Yuetong Li}
\affiliation{
  \institution{The University of Chicago}
  \city{Chicago}
  \country{USA}
}
\email{yuetong@uchicago.edu}

\author{Alex Jiang}
\affiliation{
  \institution{Shanghai High School International Division}
  \city{Shanghai}
  \country{China}
}
\email{alexjiang007@outlook.com}

\author{Dezhi Ran}
\authornote{Co-corresponding authors.}
\affiliation{
  \institution{Key Lab of HCST (PKU), MOE; SCS, Peking University}
  \city{Beijing}
  \country{China}
}
\email{dezhiran@pku.edu.cn}

\author{Yitao Hu}
\affiliation{
  \institution{Tianjin University}
  \city{Tianjin}
  \country{China}
}
\email{yitao@tju.edu.cn}

\author{Wei Yang}
\affiliation{
  \institution{University of Texas at Dallas}
  \city{Richardson}
  \country{USA}
}
\email{wei.yang@utdallas.edu}

\author{Tao Xie}
\authornotemark[1] 
\affiliation{
  \institution{Key Lab of HCST (PKU), MOE; SCS, Peking University}
  \city{Beijing}
  \country{China}
}
\email{taoxie@pku.edu.cn}

\begin{abstract} 

\thispagestyle{plain} 
\pagestyle{plain} 

To alleviate the substantial cost of manually crafting user interface (\textit{UI}) test cases, UI test migration aims to automatically generate test cases for a target mobile application (\textit{app}) by adapting those from a source app that shares similar functionalities. Traditionally, this process has been approached as a sequential UI-event-mapping problem, where events in the source app are mapped to those in the target one based on their textual descriptions. Prior research has extensively focused on enhancing the event-mapping accuracy of natural language processing (\textit{NLP}) models. Although the advent of large language models (\textit{LLMs}) with impressive NLP capabilities suggests the potential for near-perfect event-mapping, our study demonstrates that even the highly accurate event-mapping of LLMs is insufficient to address the implementation discrepancies between the source and the target apps,  reducing the overall effectiveness of LLM-driven solutions for UI test migration.

To address this challenge, in this paper, we propose \toolname{}, a skill-adaptive imitation learning framework designed to enhance the effectiveness of UI test migration through two key designs. 
First, \toolname{} leverages the source test cases as demonstrations and employs a multi-level abstraction of test cases' underlying skills, so as to extract and archive the testing information from source test cases as the knowledge base for the subsequent test generation on the target app.
Second, \toolname{}  selectively reuses a subset of the learned skills to guide the generation of test cases for the target app with its novel context- and history-aware skill adaptation. While \toolname{} can be instantiated with any imitation learning techniques, we utilize the in-context learning capabilities of LLMs, specifically GPT-4 and DeepSeek-V2, to instantiate \toolname{}. Evaluations results show that \toolname{} substantially improves the effectiveness of UI test migration, with 149\% higher success rate than state-of-the-art approaches.

\end{abstract}


\maketitle
\thispagestyle{plain} 
\pagestyle{plain} 

\section{Introduction} 
User Interface (\textit{UI}) testing has been an irreplaceable but labor-intensive approach to ensuring the quality of mobile applications (\textit{apps} for short). To alleviate the substantial cost~\cite{dobslaw2019estimating,linares2017developers,pan2020qtesting,ran2022automated} of manually crafting UI test cases, UI test migration~\cite{hu2018appflow,behrang2019test,lin2019test,zhao2020fruiter,mariani2021evolutionary,mariani2021semantic} has been proposed to migrate an existing UI test case from another app (denoted as the \textit {source app}) to the App Under Test (\textit{AUT}), denoted as the \textit{target app}, exploiting the functionality similarities~\cite{mariani2018augusto} shared across apps in the same category.
As we illustrated in Section~\ref{subsec:problem_existing}, state-of-the-art approaches formulate the UI test migration task as a UI event matching/mapping problem~\cite{liang2023rida,zhang2024learning,mariani2021semantic}.
Consequently, extensive efforts are investigated to improve the effectiveness of the mapping model, from using word embeddings~\cite{lin2019test,mariani2021semantic} to BERT~\cite{zhang2024learning} trained on migration dataset~\cite{mariani2021semantic,zhao2020fruiter}.

\begin{figure*}[t]
    \centering
    \includegraphics[width = \textwidth]{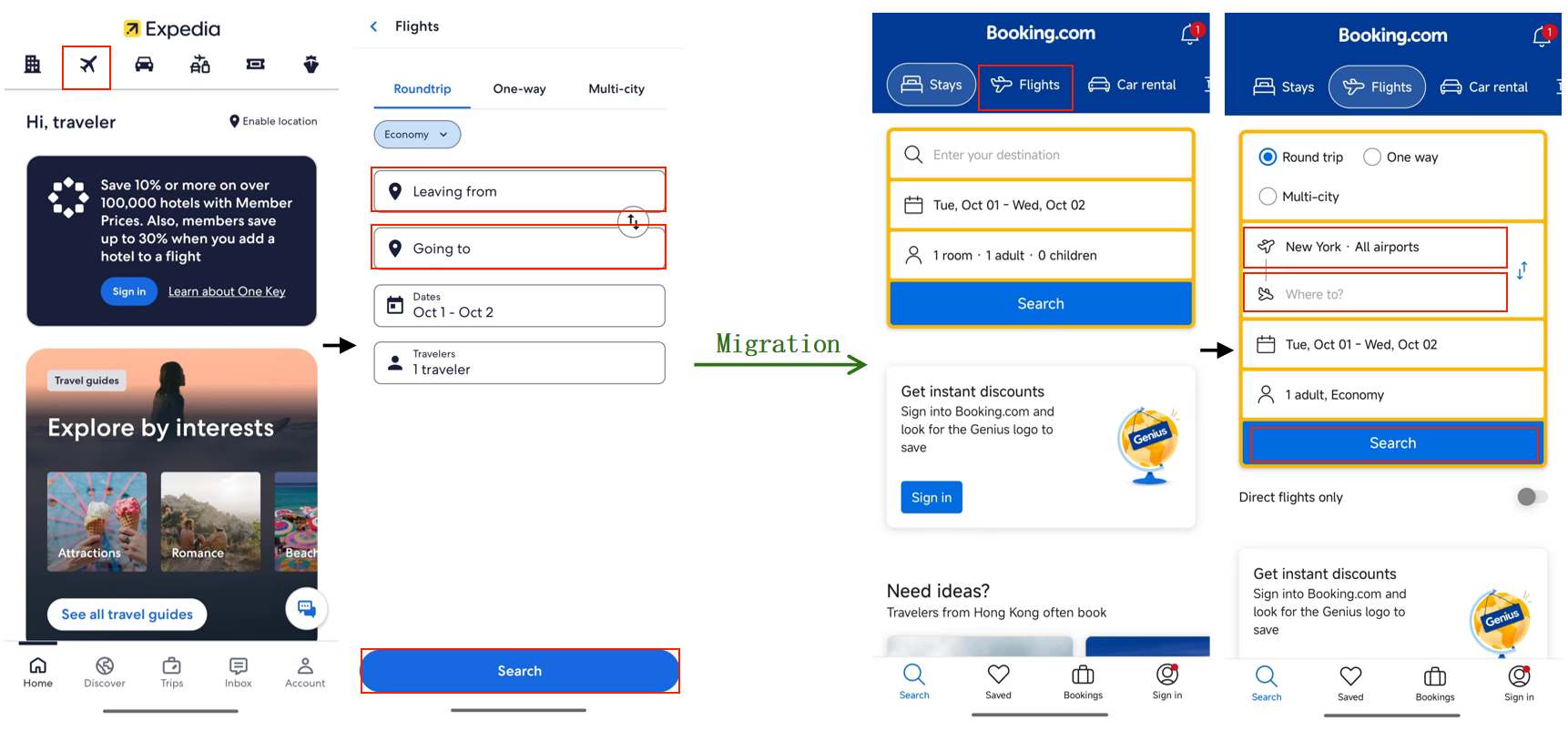}
    \caption{A direct migration from Expedia to Booking.}
    \label{fig::simple_mapping}
\end{figure*}

\begin{figure*}[t]
    \centering
  
  \begin{minipage}[b]{\linewidth}
    \subcaptionbox{ABC News}
      {\includegraphics[width=\linewidth]{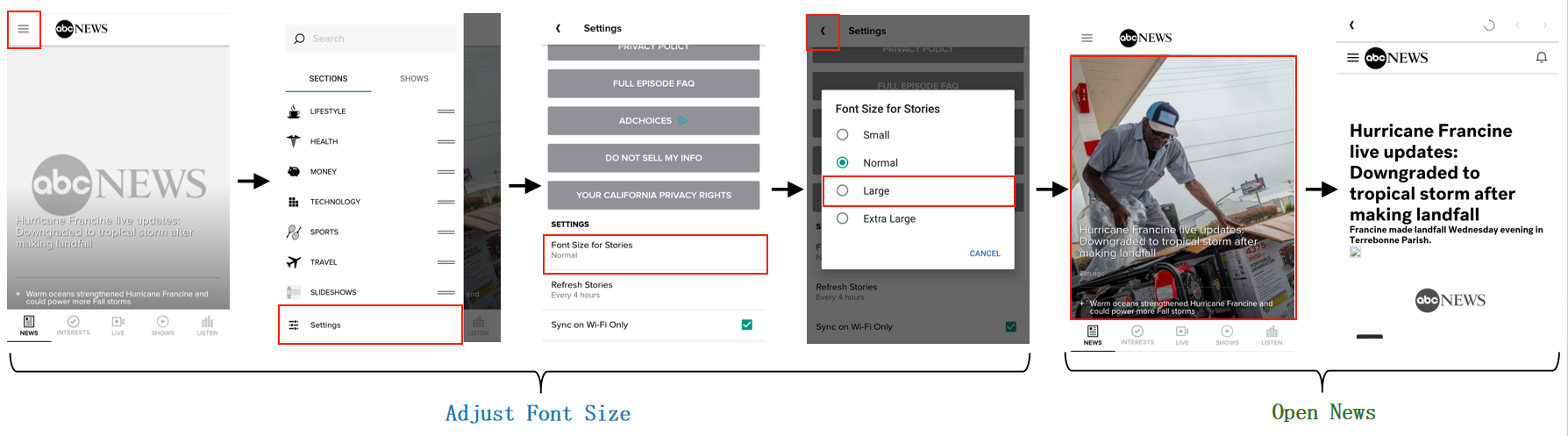}}%
  \end{minipage}%
  \hfill
  \begin{minipage}[b]{\linewidth}
    \subcaptionbox{Smart News}
      {\includegraphics[width=\linewidth]{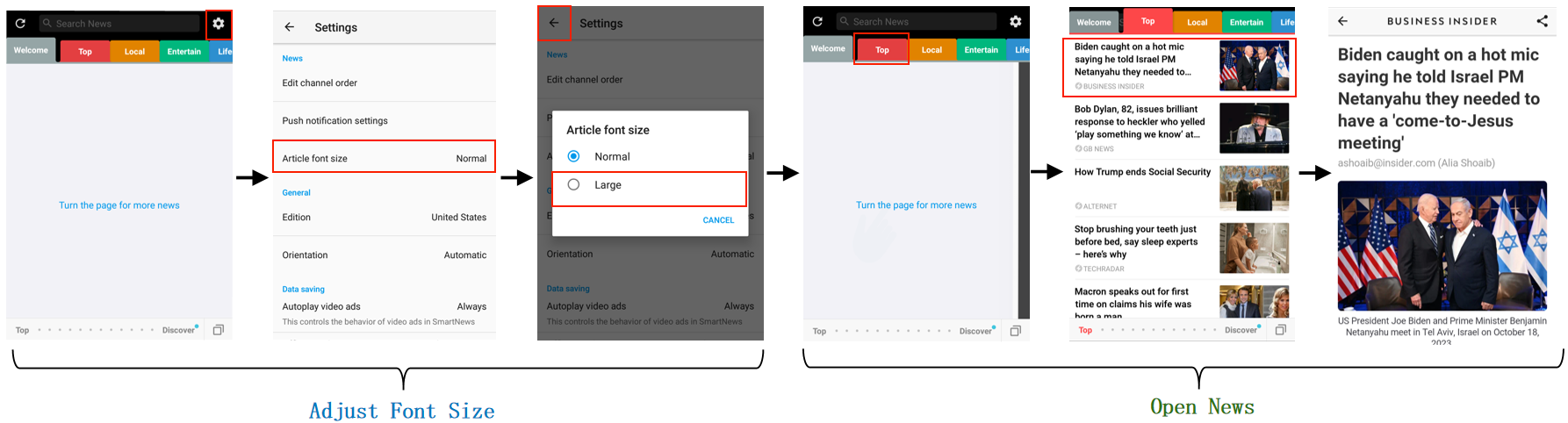}}%
  \end{minipage}%
  \hfill
  \begin{minipage}[b]{\linewidth}
    \subcaptionbox{Fox News}
      {\includegraphics[width=0.50\linewidth]{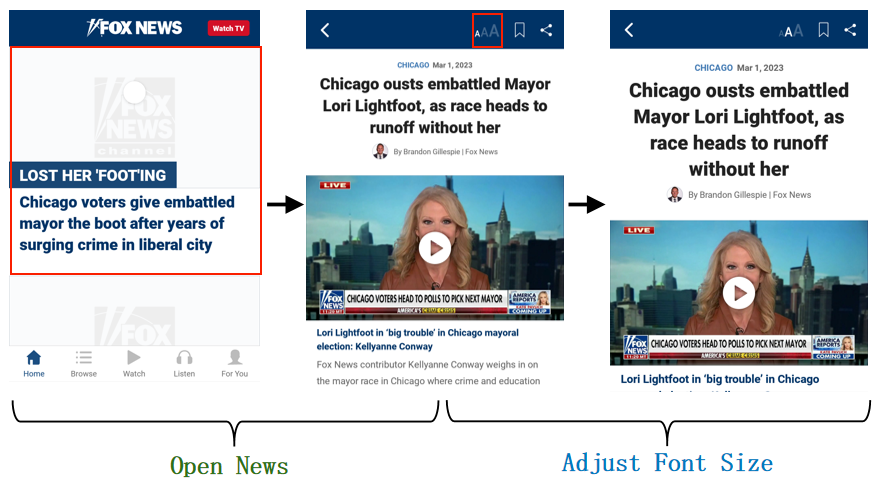}}%
    \hfill
    \subcaptionbox{Mapping from ABC News to Fox News}
      {\includegraphics[width=0.48\linewidth]
      {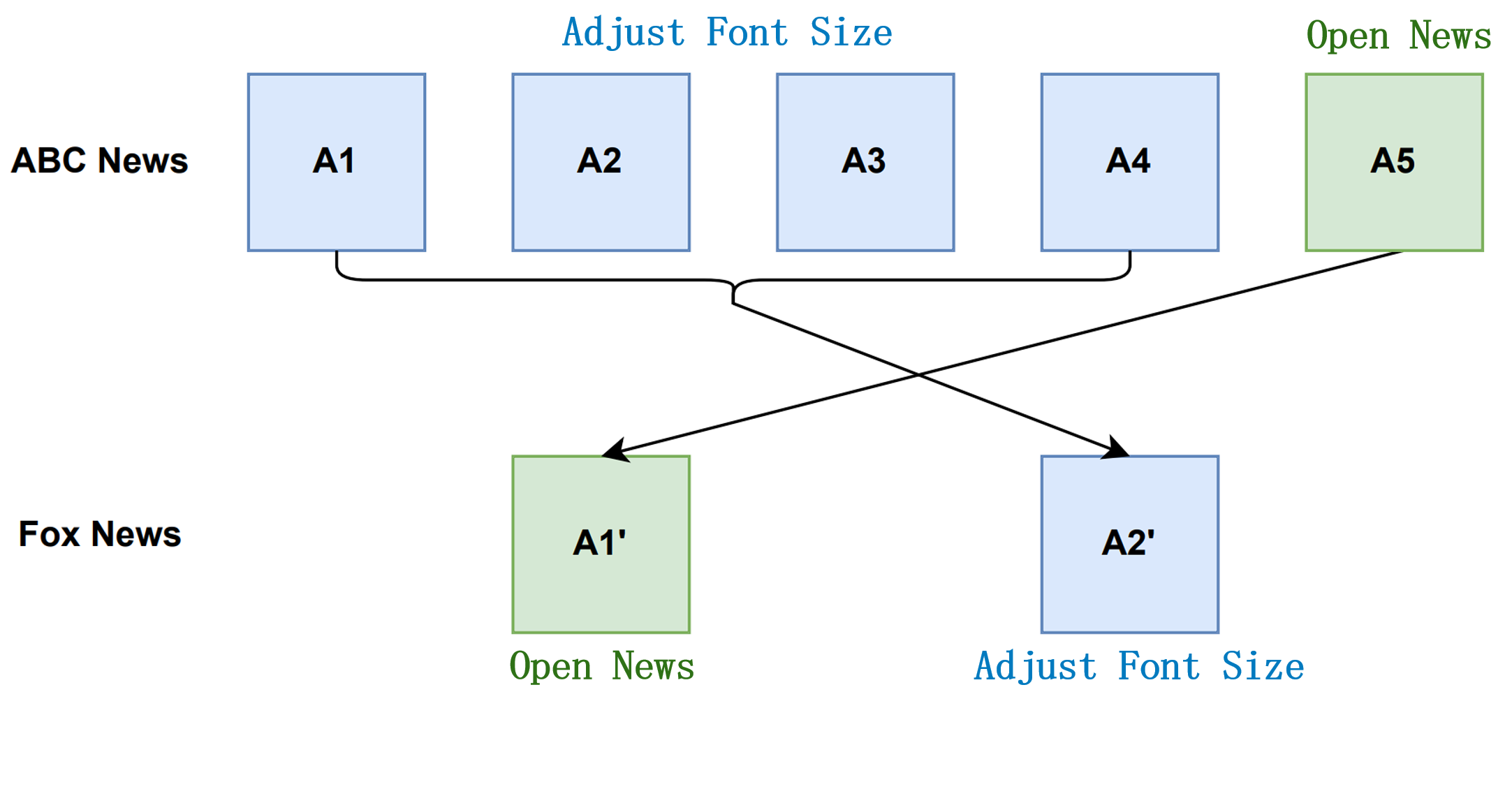}}%
  \end{minipage}
    \caption{(a)(b)(c) Three different implementations of a test case that set the font size bigger and opens a news page on ABC News, Smart News, and Fox News. The red box indicates the events that need to be selected in each UI hierarchy, and the curly braces divide the entire test case into several skills, showing the content of the skills.
    (d) The abstract mapping from ABC News to Fox News. `A\textsubscript{k}' represents the k-th event in the source test case, while `A\textsubscript{k}$'$' indicates the k-th event in the target test case. Each color represents high-level common skills shared across ABC News and Fox News.}
    \label{fig::motiv}
\end{figure*}

However, existing UI test migration approaches~\cite{liang2023rida,zhang2024learning,mariani2021semantic} assume that apps within the same category share similar UI interface styles (e.g., nearly a 1-to-1 mapping of UI elements between source and target test cases as shown in Figure~\ref{fig::simple_mapping}). Nonetheless, these approaches encounter significant limitations due to the diverse implementation strategies employed for the same functionalities (e.g., test cases in Figure~\ref{fig::motiv}). As we described in Section~\ref{subsec:problem_limitation}, even when two apps implement the same functionality, the differences in higher-level abstraction make generalizing event-mapping techniques difficult and often ineffective. Traditional event-similarity matching approaches depend heavily on a precise correspondence between UI events across different apps. As shown by our preliminary study in Section~\ref{sec::prelim}, existing state-of-the-art method CraftDroid~\cite{lin2019test} struggled to adapt to new and varied UI environments, resulting in lower performance and a lack of flexibility in handling unseen scenarios.

Given the advanced GUI comprehension abilities of Large Language Models (\textit{LLMs})~\cite{liu2022fill,feng2024prompting}, it may seem that LLMs could offer a superior solution to UI test migration with more accurate event mapping than traditional models.
LLMs' ability to understand and process complex visual and textual information suggests they would excel in achieving more precise event mapping compared to traditional models. This expectation arises from the assumption that improved comprehension directly correlates with better identification and alignment of events between different user interfaces.
However, our preliminary investigation  reveals two important insights. While LLMs did demonstrate high accuracy in event mapping (Section~\ref{subsec:prelim_rq1}), confirming their proficiency in understanding GUI content, they failed to achieve high effectiveness in UI test migration (Section~\ref{subsec:prelim_rq2}). 
The experiments indicated that event matching was not the main challenge; instead, the existing workflow itself becomes the bottleneck.
For instance, as shown in Figure~\ref{fig::motiv}(d), when the mapping involves reversed order and abstraction from the source test case, LLMs are limited in their ability to effectively handle the migration task without external adaptations or restructured workflows.

\noindent\textbf{Solution.} 
To address the limitation of existing workflows in UI test migration, we realized that the core issue lies in the rigidity of current approaches when dealing with complex scenarios. 
Our insight contains two aspects: (1) Migration tool must have a better understanding to the functionality of source test case, rather than directly calculate the similarity of each event. (2) Migration tool must have a more flexible framework to generate the event, to be capable of adapting to diverse and dynamically changing environments.
Based on this insight, we reformulate UI test migration with \toolname{}, a \textit{Skill-Adaptive Imitation Learning} (\toolname{}) framework.
The reformulation of the UI test problem is detailed in Section~\ref{subsec:problem_reformulation}, and the approach is thoroughly explained in Section~\ref{sec::approach}.
\toolname{} is designed to learn from the behaviors in a source test case to achieve similar goals in different environments. It decomposes complex expert behaviors into subtasks, referred to as skills.
The framework consists of two designs: a high-level skill learner and a lower-level event learner.
The skill learner focuses on high-level decision-making and strategy planning, identifying which skills are needed to achieve the overall goal.
The event learner then selects the events required to complete these skills in the given environment.

While the preceding SAIL framework can be implemented with any imitation learning algorithm, to bypass the cumbersome training costs, we exploit the LLM's in-context learning capability and verbal imitation learning to instantiate \toolname{} with two major designs.
(1) \textit{LLMs as High-Level Learners:} At the high level, LLMs are tasked with understanding the broader objectives and strategic requirements of UI tests. 
By interpreting test cases into disjoint and functionally independent skills, LLMs identify which skill of a source test case can be functionally applicable to the target app. 
The high-level learner thus leverages the source test cases as demonstrations and employs the underlying skills to ensure that the essential goals of UI testing are preserved and accurately migrated, even when direct event matching are lacking.
(2) \textit{LLMs as Low-Level Learners:} At the low-level, LLMs take the skills as in-context examples to address the execution specifics of the aligned test cases, adapting them to the nuances of the target apps by recognizing and accommodating minor differences in how similar skills are implemented. 
In this way, the low-level learner reuse a subset of the learned skills from the source test case, translate the high-level functional skills into executable test events on the target app.

To evaluate the effectiveness of \toolname{}, we instantiate \toolname{} with three LLMs: GPT-4o, GPT-4o-mini and DeepSeek-V2. We then compare them against two state-of-the-art Android UI migration tools, CraftDroid~\cite{lin2019test} and SemFinder~\cite{mariani2021semantic}, on \testNum{} test cases from \appNum{} popular apps used by existing work~\cite{lin2019test,zhao2020fruiter,mariani2021semantic}.
Evaluation results show that \toolname{} substantially outperforms state-of-the-art test migration tools, evidencing a 149\% increase in migration success rate across the entire dataset and a notable 294\% increase in instances where the source and target test cases lacked direct 1-to-1 mappings.
We further conduct extensive experiments on individual algorithms of \toolname{}, and experimental studies reaffirm the potency of the imitation learning formulation and the aptitude of prompt engineering in embodying \toolname{}.

In summary, this paper makes the following main contributions:
\begin{itemize}
\item  \toolname{}, a skill-adaptive imitation learning framework for reformulating UI test migration as an imitation learning problem.
\item A training-free instantiation of \toolname{} with LLMs.
\item Extensive evaluations demonstrating the efficacy of \toolname{} against state-of-the-art approaches.
\end{itemize}
\section{Problem Formulation}\label{sec::problem}
In this section, we first discussed the definition of UI test migration, summarized the workflow of existing work, and then pointed out the limitations faced by existing work. Finally, we provided our reformulation of UI test migration based on imitation learning.

\subsection{UI Test Migration}\label{subsec:problem_existing}

UI test migration involves transferring an existing UI test case from a source app to a target app to reduce manual testing efforts. This process takes the source test case as input and focuses on replicating the sequence of UI events that represent user interactions with UI elements. These elements are identified by properties such as resource IDs and content descriptions. During migration, the process iteratively executes UI events on the AUT and observes its responses to gather essential information, such as the UI hierarchy and activity, and then extract available events in current UI screen to select one. The goal is to generate an equivalent event sequence on the AUT that mirrors the functionality and intent of the original test case.

Existing approaches~\cite{behrang2019test,lin2019test,mariani2021evolutionary,zhang2024learning} to UI test migration formulate the task as a UI event matching problem focus on mapping UI event sequences from a source app to corresponding events in a target app. This involves a repetitive process: starting with the first UI event from the source app, a language model like BERT calculates its similarity to each available event on the current UI screen of the target app, using UI properties such as resource ID and content description. The target app event with the highest similarity is selected and executed. This process is repeated for each subsequent event until all events from the source app are processed. Additionally, special cases are managed by maintaining UI state transition graphs or addressing mismatches when similarity is too low. For instance, Figure~\ref{fig::simple_mapping} shows a direct 1-to-1 mapping from the Expedia app to the Booking app. The initial event of clicking on the flight icon in Expedia is matched to the most similar event in Booking, which is clicking on the flight tab. This matching and execution process continues with each event from Expedia until all events are mapped to Booking.

\subsection{Limitation of Existing Work}\label{subsec:problem_limitation}

Due to the reliance on event matching, existing methods can migrate the test case in Figure~\ref{fig::simple_mapping} easily. However, real-world migrations are often complex. Even if two apps in the same category share similar functionalities, their specific implementations can vary significantly. A complete 1-to-1 mapping from source test cases to target test cases is frequently unavailable. Existing approaches primarily depend on event matching, which heavily relies on UI events to correspond accurately from the source app to the target app. This focus often overlooks the fundamental goal of migration: to replicate the functionality and intent of the original test case. Consequently, diverse implementation strategies for the same functionalities pose a significant limitation for existing methods, which we will analyze in detail in Section \ref{subsec:prelim_rq2}.

For example, Figure~\ref{fig::motiv} illustrates three versions of the same test on different apps, showing the challenges of migrating from ABC News to Smart News (\(T_{a\rightarrow b}\)) and the near impossibility of migrating to Fox News \(T_{a\rightarrow c}\).
The test adjusts the font size and opens a news page to show the effect of the font change. The test is performed on ABC News, Smart News, and Fox News, corresponding to Figure~\ref{fig::motiv}(a)-(c). In the test, a complete 1-to-1 mapping from the source app to the target app is absent.

Traditional approaches struggle with migrating \(T_{a\rightarrow b}\).
In ABC News, accessing the settings involves clicking a menu, while Smart News offers direct access but requires an extra step to open a news page. These differences make sequential event mapping approaches ineffective for \(T_{a\rightarrow b}\), as key actions in ABC News have no equivalents in Smart News.

Migrating from ABC News to Fox News (\(T_{a\rightarrow c}\)) is nearly impossible because it requires reversing the order of events---Fox News places the font change button on the news page rather than in settings, so the test must first open a news page to adjust the font size.

Our experiment results on CraftDroid support our analysis.
Although previously considered state-of-the-art, CraftDroid only manages to click headline news on \(T_{a\rightarrow b}\). Even worse, on \(T_{a \rightarrow c}\), CraftDroid incorrectly navigates to an unrelated page, failing to generate any valid result.

\subsection{Problem Reformulation} \label{subsec:problem_reformulation}

To address the preceding limitation of existing workflow, we reformulate the problem of UI test migration based on imitation learning, and propose \toolname{} framework.

Imitation learning~\cite{duan2017one} is a general approach, in which the goal of the learner is to imitate the behavior of the expert on a task.
Formally speaking, an agent $\mathcal{A}$ (learning algorithm) is trained to learn from a set of expert demonstrations where each demonstration is defined as a set of observation/action pairs $\xi = \{(x_i,a_i)|i\}$.
The goal of the learning is to produce a policy mapping $\hat{\pi}*$ between observations $x$ and actions $a$ from the demonstrations given by the expert mapping $\pi*$. 

We now consider a UI test migration problem.
Under this setting, a test migration tool $\mathcal{A}$ is trying to migrate a test case $\xi$ comprising of a sequence of UI screen/action pairs $\{(x_i, a_i)|i\}$ onto new apps.
The goal of the UI test migration is to generate a test case that resembles the implementation of the source test case.
Since the source test case is a manifest of the expert's policy function $\pi*$, $\mathcal{A}$ needs to learn from the source test case $\xi$ a policy function $\hat{\pi}*$ that is similar to $\pi*$, successfully imitating the behavior.
In order to achieve this, the test migration tool should be able to learn from the demonstration the intents of each step of the test sequence, which can be summarized as skills.

Based on this formulation with imitation learning, we propose \toolname{}. The \toolname{} framework, as detail in Section~\ref{sec::approach}, can migrate the test case in Figure~\ref{fig::motiv} easily. As is shown in Figure~\ref{fig::motiv}, each of the test cases can be divided into sets of skills (represented by different curly braces).
Thus, after \toolname{} learns the skills needed from the source test, it can replay the skills on the target app.
When performing $T_{a\rightarrow c}$, Figure~\ref{fig::motiv}(d) shows the abstract mapping.
\toolname{} first learn the two skills ``Setting Font'' and ``Open News''.
Next, \toolname{} identifies that the skill ``Setting Font'' is not available in initial UI screen and opt to open the news first, after which it adjusts the font.
\section{Preliminary Study}\label{sec::prelim}
To support our motivation described in Section~\ref{sec::problem}, we conduct a preliminary study to show the defect of existing migration workflow.
Existing UI test migration approaches heavily focus on improving the effectiveness of a mapping model, which maps UI elements from the source test case to the UI elements in the target app based on textual information.
Given the superior effectiveness of LLMs in software engineering~\cite{codex, liu2022fill} and testing tasks~\cite{wang2023enabling, pan2021empirical, schafer2023adaptive}, we expect that using LLMs to replace the mapping models can resolve the challenge of UI test migration, i.e., developing a ``gold'' mapping model.
However, our evaluation in Section~\ref{subsec:prelim_rq2} shows that replacing the mapping model of an available state-of-the-art approach CraftDroid~\cite{lin2019test} with an LLM exhibits low effectiveness when applied to UI test migration.
This observation leads us to investigate the causes of the low effectiveness of LLM-based UI UI test migration.
Specifically, we hypothesize that (1) state-of-the-art LLMs may not be capable of mapping UI elements based on their natural language descriptions compared to deep-learning models trained on specific UI data, or (2) existing workflow of UI test migration limits the power of LLMs.

To find the answer to the preceding hypothesis, we conduct a preliminary study aimed at answering the following research questions:

\begin{itemize}
    \item \textbf{RQ1}: How effectively can LLMs map UI events from the source app to the target app?
    \item \textbf{RQ2}: How effectively can LLMs migrate UI test cases with existing workflow?
\end{itemize}

\subsection{RQ1: Zero-Shot UI Event Matching Capability of LLMs} \label{subsec:prelim_rq1}

Previous research in UI test migration has focused on developing optimal event matching models. 
Event matching aims to find a best match for each of the events in the source test case in the target app in terms of similarity, etc.
LLMs have demonstrated remarkable zero-shot natural language understanding capabilities across a wide range of tasks. 
This section aims to investigate whether LLMs could be the definitive solution for this challenge.

\noindent\textbf{Experiment Setup}.
For event matching, we utilized the SemFinder dataset~\cite{semfinder-dataset}, which is specifically tailored for this task. The SemFinder dataset includes popular apps with more than 1,000 downloads from Google Play~\cite{google-play} and provides event matching relations, comprising 295 should-be-matched and 4,649 should-not-be-matched event pairs. This dataset has been used in previous research~\cite{zhang2024learning}.

However, we identified several limitations in the SemFinder dataset. We found that its ground truth annotations were not as accurate as initially assumed. Additionally, the should-not-be-matched event pairs often included events from different screens within the target app. Since different screens may share events with similar functions, this could lead to unnecessary confusion for matching models.

To address these issues, we used the dataset for migration as described in Section~\ref{sec::setup} and annotated the mapping from source test case events to target test case events. 
For each event in the source test case, we identified the corresponding UI hierarchy in the target app using this mapping and collected all available events in the current UI hierarchy.
We observe that a large number of UI screens wrap multiple UI elements providing information regarding description and padding around the actually interactable one, causing ambiguity when selecting the events and judging for the correctness.
Thus, to ensure the fairness when determining the correctness of an event match, we labeled an event as correct if the center of the widget in the event fell within the bounds of the ground truth event in the target app test case. 
Our dataset provides 3,765 should-be-matched and 62,699 should-not-be-matched event pairs, offering a clearer illustration of the event matching capabilities of various models.

\noindent\textbf{Evaluation Metrics}.
For the SemFinder dataset, we employed two widely recognized statistical metrics for matching tasks: Mean Reciprocal Rank (MRR) and Top-1 accuracy, as adopted in previous studies~\cite{mariani2021semantic}. 
For our dataset, we exclusively used the Top-1 metric, as it is the most critical measure in event matching. 

MRR is the average of the reciprocal ranks of the total queries \( Q \). It provides a measure of how well the model ranks the correct event pairs across all queries. The formula for MRR is given by:

\begin{equation}
    MRR = \dfrac{1}{|Q|}\sum\limits_{i=1}^{|Q|}\dfrac{1}{rank_i}
\end{equation}

Top-1 accuracy, on the other hand, is the ratio of queries where the ground truth event has the highest similarity score in the returned list of event pairs. The formula for Top-1 accuracy is:

\begin{equation}
    Top1 = \dfrac{1}{|Q|}\sum\limits_{i=1}^{|Q|}
    \begin{cases}
    1, & \text{if $rank_i=1$}\\
    0, & \text{otherwise}
    \end{cases}
\end{equation}

\noindent\textbf{Baselines}.
For the SemFinder dataset, we utilized several established approaches as baselines. Specifically, the CraftDroid and SemFinder are natively implemented within the SemFinder source code. Additionally, we applied various LLMs to the SemFinder dataset to evaluate their performance.

For our dataset, we implemented the CraftDroid approach, the SemFinder approach, and the LLM approach using different LLMs. Our dataset includes app screenshots, enabling us to also implement Visual Language Models (VLMs) to assess their performance.

According to the report in~\cite{zhang2024learning}, TEMDroid is the state-of-the-art model for event matching task. However, the model parameters of TEMDroid is private and source code of TEMDroid is not ready for reproduce. That is why we didn't contain TEMDroid as our baselines.

\noindent\textbf{Study results}.
In both the SemFinder Dataset and Our Dataset, all methods were evaluated under a zero-shot scenario, as shown in Table~\ref{tab::rq1}. The Top1 accuracy results underscore the adaptability of various approaches to cross-app UI event matching tasks.

On the SemFinder Dataset, LLM-based methods exhibited strong performance. DeepSeek-V2, GPT-4o and GPT-4o-mini achieved Top1 accuracies of 73.9\%, 73.9\%, and 73.6\%, respectively. Although not trained on this dataset, these models demonstrated strong generalization, highlighting the robustness of LLMs in understanding UI elements and event matching.

In contrast, non-LLM methods, such as SemFinder and CraftDroid, showed lower performance. SemFinder achieved a Top1 accuracy of 67.1\%, but CraftDroid, at 48.4\%, lagged significantly. CraftDroid relies on the Word2Vec~\cite{mikolov2013distributed} model, using predefined rules for UI event matching. While Word2Vec captures word semantics, it lacks the capacity of LLMs to comprehend complex UI structures and contexts, leading to inferior performance.

On Our Dataset, DeepSeek-V2 results in an impressive Top1 accuracy of 88.2\%, showing its adaptability to new UI structures without task-specific training. GPT-4o followed with a Top1 accuracy of 86.9\%, again demonstrating the effectiveness of LLMs in zero-shot scenarios.

However, Visual Language Models (VLMs) faced difficulties with matching tasks. GPT-4o-Vision and GPT-4o-mini-Vision achieved Top-1 accuracies of 42.7\% and 29.5\%, respectively. This is due to challenges in visual grounding, as VLMs depend on screenshots that often lack enough semantic information, making it difficult to recognize UI elements and interactive regions. Large models typically find it easier to analyze text rather than corresponding screenshots, particularly for tasks like identifying a button with a specific text label.

Traditional non-LLM methods showed continued limitations on Our Dataset. CraftDroid and SemFinder achieved Top1 accuracies of 56\% and 54.4\%, respectively, highlighting their dependence on predefined rules and limited model capabilities. This restricts their adaptability to diverse UI elements compared to LLMs.

\noindent\fcolorbox{black}{gray!20}{
\begin{minipage}{\dimexpr\linewidth-2\fboxrule-2\fboxsep\relax}
\Answer{RQ1}{{
These findings highlight LLMs' strong generalization abilities, making them more effective than traditional models for cross-app UI event mapping. Thus, event matching is not the bottleneck for UI test migration, as initially hypothesized. LLMs have demonstrated their capability to perform event matching in zero-shot settings. }}
\end{minipage}
}

\begin{table}[t]
\centering
\caption{Metrics of different event matching approaches.}
\label{tab::rq1}
\renewcommand{\arraystretch}{1.2}
\begin{threeparttable}
\begin{tabular}{lccc}
\toprule
\multirow{2}{*}{\textbf{Approach}} & \multicolumn{2}{c}{\textbf{SemFinder Dataset}} & {\textbf{Our Dataset}} \\
\cmidrule(lr){2-3}
\cmidrule(lr){4-4}
 & \textbf{Top1} & \textbf{MRR} & \textbf{Top1} \\
\midrule
\textbf{LLMs/VLMs} & & &\\
\cmidrule(r){1-1}
DeepSeek-V2 & \textbf{73.9} & \textbf{85.1} & \textbf{88.2}\\
GPT-4o & \textbf{73.9} & 84.9 & 86.9\\
GPT-4o-mini & 73.6 & 83.6 & 79.2\\
GPT-4o-Vision & / & / & 42.7\\
GPT-4o-mini-Vision & / & / & 29.5\\

\midrule
\textbf{non-LLM} & & & \\
\cmidrule(r){1-1}
Craftdroid\cite{lin2019test}& 48.4 & 67.6 & 56.0\\
SemFinder\cite{mariani2021semantic}& 67.1 & 79.6 & 54.4\\
\bottomrule
\end{tabular}
\begin{tablenotes}
\small
\item \textit{Note: As detailed in Section \ref{subsec:prelim_rq1}, the results denoted by `/' could not be obtained.}
\end{tablenotes}
\end{threeparttable}
\end{table}

\subsection{RQ2: Zero-Shot UI Test Migration Capability of LLMs}
\label{subsec:prelim_rq2}
Building on the findings from Section \ref{subsec:prelim_rq1}, which demonstrate the strong UI event matching capabilities of LLMs, we now investigate whether LLMs can effectively handle the UI test migration task. This section evaluates the test cases migrated by LLMs and existing approaches, highlighting the major failure reasons for both.

\noindent\textbf{Experiment Setup.}
The dataset of migration, evaluation metric, test platform is the same as Section ~\ref{sec::setup}.

\noindent\textbf{Baselines.}
To evaluate the UI test migration capabilities of LLMs, we integrated our LLM-based approach into the existing CraftDroid framework~\cite{lin2019test}, replacing its original matching module. We then compared the performance outcomes.
In RQ2, we contain CraftDroid as baseline since it is the state-of-the-art approach in the tools with source code and model parameters public that we can reproduce in our dataset.

\noindent\textbf{Study Results}.
The results in Table \ref{tab::rq3} demonstrate the UI test migration capabilities of the baseline methods. The original CraftDroid achieves a success rate of 25.5\%. When the matching module is replaced with a large language model, CraftDroid-GPT-4o-mini improves to 27.3\%. Additionally, CraftDroid-DeepSeek-V2 reaches 28.8\%.
These findings suggest that the UI test migration capabilities of existing methods do not significantly benefit from the advanced event matching capabilities of LLMs. Additionally, we observe that when there is a 1-to-1 mapping from the source to the target test case event sequence, the success rate is significantly higher—approximately three times greater than in non-1-to-1 mapping scenarios.

Upon careful examination of the test cases generated by the baseline methods, we identified three primary reasons for failure in the existing workflows, as highlighted by the motivating example in Figure~\ref{fig::motiv}(a)-(c):

\textbf{Extra Event in the Source Test Case.}
When there is an extra event in the source test case, the migration tool must decide whether to ignore it, which presents a challenge. Misjudging an extra event can lead to navigation to an incorrect UI state, complicating the achievement of the test objective.

\textbf{Missing Event in the Source Test Case.}
The issue of missing events is symmetrical to that of extra events but poses a more significant challenge for migration tools like CraftDroid. The tool needs to explore the UI space to identify the matching event. While CraftDroid calculates button similarity to guide exploration, it lacks an understanding of transitions between UI states.

\textbf{Reverse Ordered Events.}
The event sequence in the source test case and the target test case may be reversed. These reverse-ordered events hinder existing sequential matching-based approaches from generating a complete test case.

\noindent\fcolorbox{black}{gray!20}{
\begin{minipage}{\dimexpr\linewidth-2\fboxrule-2\fboxsep\relax}\Answer{RQ2}{
Despite the strong UI event matching capabilities of LLMs demonstrated in RQ1, they cannot directly solve the UI test migration task using the workflow of existing approaches. The current workflow struggles in complex scenarios where there is not a 1-to-1 sequential mapping between the source and target test cases, limits the power of LLMs.
}
\end{minipage}
}
\section{\toolname{} Approach} \label{sec::approach}

\begin{figure*}[ht]
    \centering
    \includegraphics[width = \textwidth]{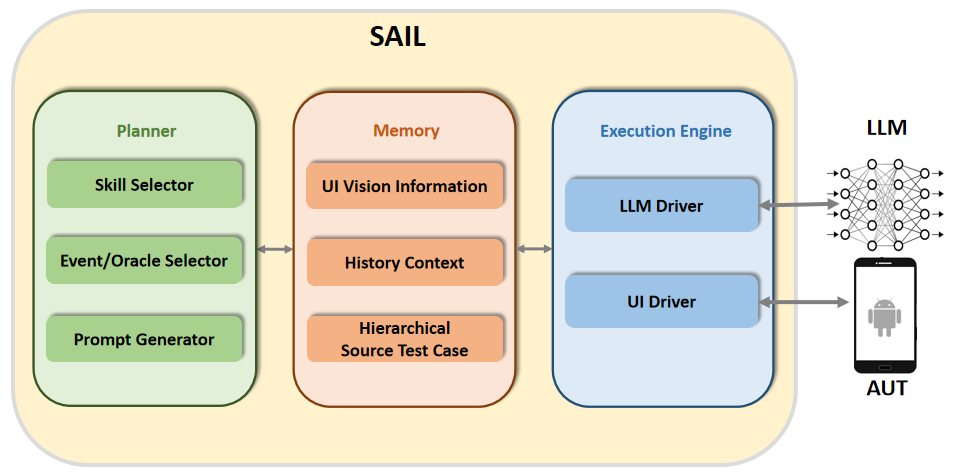}
    \caption{System architecture of \toolname{}.}
    \label{fig::system}
\end{figure*}

\subsection{System Architecture of \toolname{}}

Figure~\ref{fig::system} presents the system architecture of \toolname{}.
\toolname{} takes the source test case, an AUT, and an LLM as inputs.
To improve the effectiveness of UI test migration, \toolname{} consists of three modules:  \textit{Execution Engine}, \textit{Memory Module} and \textit{Planner}.

\subsubsection{Execution Engine.}

The execution engine in the \toolname{} plays a critical role in interfacing with both the LLM and the AUT. It comprises two main components:

\noindent \textbf{LLM Driver}.
The LLM driver is responsible for interacting with the LLM. It sends prompts to LLM. It receives the raw response from LLM, parses the response of LLM to the format we need(e.g. Yes or No, an UI Element), which can be used in \toolname{}.

\noindent \textbf{UI Driver}.
The UI driver manages the interaction between \toolname{} and AUT. For an event generated in \toolname{}, the UI driver perform it on AUT. Also, the UI driver dumps the UI hierarchy file and screenshot representing the current screen on AUT. Based on the raw UI hierarhcy, the UI driver perfrom similar UI aggregation to CraftDroid\cite{lin2019test} and extract interactable UI elements. Based on the screenshot, the UI driver add vision information to buttons without any description on UI hierarchy.

\subsubsection{Memory Module.}

The memory module in \toolname{} system architecture plays a crucial role in storing and managing various types of information necessary for \toolname{}. It is divided into three main components:

\noindent \textbf{UI Vision Information.}
We use GPT-4o to incorporate UI Vision information, storing it in the memory module for retrieval.
Previous studies\cite{li2022spotlight, baechler2024screenai} have highlighted the absence of essential descriptions (e.g., content-desc attributes) in critical UI elements, such as buttons. To address this deficiency, we developed the UI Vision Information Module, which leverages both a screenshot and the UI hierarchy as inputs. For each element requiring a description, the module captures the specific region of the screenshot where the element is located and then utilizes GPT-4o to analyze this region in conjunction with the original screenshot, thereby extracting the visual information pertinent to the element.
Upon extracting the visual information, the module stores this data along with a hash of the element's image. In subsequent instances when the vision information of this UI element is required, the UI Vision Information Module can simply hash the element and retrieve the stored visual information using the hash, thereby obviating the need to query the LLM again.
Our experiments demonstrate that this method of obtaining UI vision information is cost-effective, accounting for only 4\% of token usage within the \toolname{} workflow.

\noindent \textbf{Historical Context.}
The historical context maintains a comprehensive record of all UI contexts encountered during the exploration of the AUT. Specifically, it encompasses the planned UI events generated by the LLM, the UI hierarchies, and the activity names observed throughout the exploration of the AUT.

\noindent \textbf{Hierarchical Source Test Case.}
When stored in \toolname{}'s memory, we organize the source test case into a three-level hierarchical tree structure, which is initially a sequence of events executed on the source app. At the highest level is the goal, representing the test objective of the test case. The goal encompasses various skills required to achieve it, which constitute the middle level. Each skill, in turn, consists of a series of sequential events from the source test case, forming the lowest level. The detailed process of generating this hierarchical structure is elaborated in Section \ref{sec::workflow}.

\subsubsection{Planner.}

The planner module of \toolname{} is comprised of three sub-modules: the skill selector, the event selector, and the prompt generator.

\noindent \textbf{Skill Selector.} The skill selector selects relevant skill from the source test case as in-context example based on the current UI screen of the AUT. When an event is performed on the AUT, the skill selector utilizes the history context to determine whether a skill has been completed. The detailed workings workflow of the skill selector are elaborated in Section \ref{sec::workflow}.

\noindent \textbf{Event Selector.} Given a list of potential events that can be performed on the AUT, the event selector first invokes the skill selector to retrieve a relevant skill as in-context example. It then prompts the LLM to select an appropriate event based on the retrieved skill.

\noindent \textbf{Prompt Generator.} The prompt generator uses a templated approach to generate prompts. The other modules pass parameters to the prompt generator, which then produces the required prompt. The design and implementation of these prompts are discussed in detail in Section \ref{sec::prompt}.

\begin{figure*}[t]
    \centering
    \includegraphics[width = \textwidth]{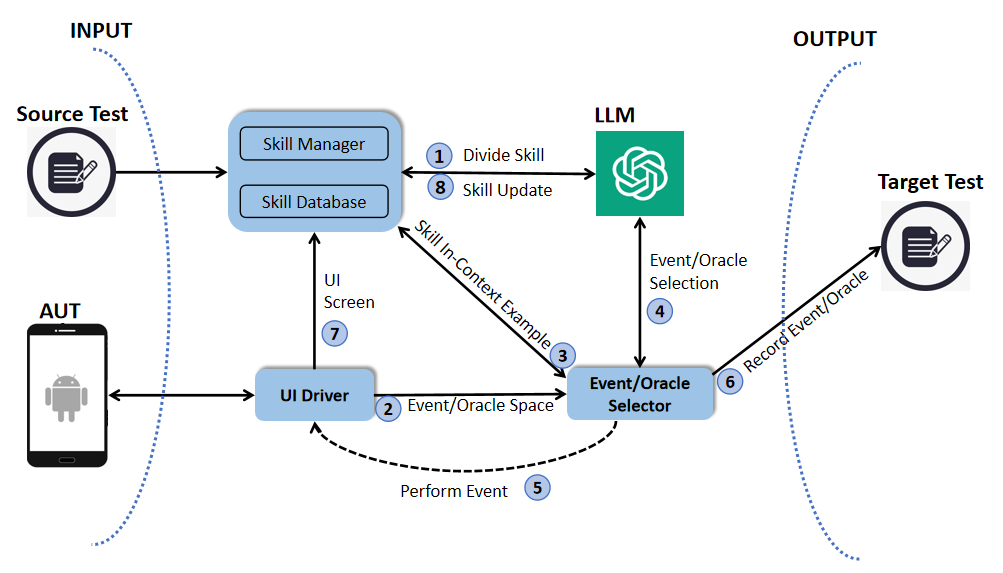}
    \caption{Workflow of \toolname{}. \toolname{} first divide task in Step 1, then repeats Step 2 to Step 8 till LLM reports the test objective reached.}
    \label{fig::workflow}
\end{figure*}

\begin{algorithm}[t]  
  \small
  \caption{Main algorithm of \toolname{}.}  
  \label{alg::main}  
  \begin{algorithmic}[1]
    \Function{\toolname{}}{$\mathtt{source\_test,\ AUT\_driver}$}
        \State{$\mathtt{goal \gets \Call{GoalConclude}{source\_test}}$}
        \State{$\mathtt{skills \gets \Call{SkillDivide}{source\_test,\ goal}}$}\algorithmiccomment{Step 1}
        \State{$\mathtt{history \gets []}$}
        \State{$\mathtt{ui\_hierarchy \gets AUT\_driver.dump\_hierarchy()}$}
        \While{\textbf{not} $\mathtt{\Call{GoalFinished}{goal,\ history,\ ui\_hierarchy}}$}\label{alg::finished_goal}
            \State{$\mathtt{available\_events \gets ui\_hierarchy.events()}$}\algorithmiccomment{Step 2}\label{alg::loop_start}
            \State{$\mathtt{relevant\_skill \gets \Call{RetrieveSkill}{skills,\ ui\_hierarchy,\ available\_events}}$}\algorithmiccomment{Step 3}\label{alg::skill_select}
            \State{$\mathtt{selected\_event \gets \Call{EventSelect}{relevant\_skill,\ available\_events}}$}\algorithmiccomment{Step 4}\label{alg::event_select}
            \State{$\mathtt{AUT\_driver.perform(selected\_event)}$}\algorithmiccomment{Step 5}
            \State{$\mathtt{history.append(selected\_event)}$}\algorithmiccomment{Step 6}
            \State{$\mathtt{ui\_hierarchy \gets AUT\_driver.dump\_hierarchy()}$}\algorithmiccomment{Step 7}
            \If{$\mathtt{\Call{SkillFinished}{relevant\_skill,\ goal,\ history,\ ui\_hierarchy}}$}\label{alg::finished_skill}
                \State{$\mathtt{skills.remove(relevant\_skill)}$}\algorithmiccomment{Step 8}\label{alg::loop_end}
            \EndIf
        \EndWhile
        \State{\Return{$\mathtt{history}$}}
    \EndFunction
  \end{algorithmic}  
  \begin{flushleft}
    \textbf{Note:} The pseudo code uses the event selector to illustrate the workflow. The oracle selector follows the same workflow as the event selector.
  \end{flushleft}
\end{algorithm}

\subsection{Workflow of \toolname{}}

\subsubsection{Workflow Overview.} \label{sec::workflow}

Figure~\ref{fig::workflow} and Algorithm~\ref{alg::main} illustrate the workflow of \toolname{} for prompting the LLM to generate UI events or oracles. As the oracle generation follows the same process as the event selector, we will focus solely on the event generation workflow.

\toolname{} begins with the input of a source test and an AUT driver. Initially, the skill manager determines the goal of the source test, guiding the LLM to understand the task at a high level and subsequently dividing the test case into multiple skills, as outlined in Step 1.

Following this, \toolname{} iterates through Steps 2 to 8 (Lines~\ref{alg::loop_start} to \ref{alg::loop_end} in Algorithm~\ref{alg::main}) until the LLM reports that the test goal has been reached (Line~\ref{alg::finished_goal} in Algorithm~\ref{alg::main}). In Step 2, the UI driver generates all available events based on the UI hierarchy of the AUT. This UI hierarchy, along with the available events, is then passed to the Event Selector.

In Step 3, the Event Selector retrieves the relevant skill from the Skill Database based on the current UI hierarchy and available events, ensuring the selection of the most appropriate skill. Subsequently, in Step 4, the Event Selector chooses an event from the available events using the retrieved skill. This event selection is guided by the skill and the current context of the UI. 
Once an event is selected, the AUT driver performs the event (Step 5), and the chosen event is recorded in the history (Step 6). \toolname{} then checks if the skill has been completed by evaluating the updated history and current UI hierarchy. If a skill is completed, it is removed from the Skill Database (Step 8).

This comprehensive workflow ensures systematic and efficient test generation, leveraging the interplay between skill management, UI interaction, and event selection to achieve the desired testing objective.

\subsubsection{Prompt and Parser Design.} \label{sec::prompt}

\begin{figure*}[ht]
    \centering
    \includegraphics[width = \textwidth]{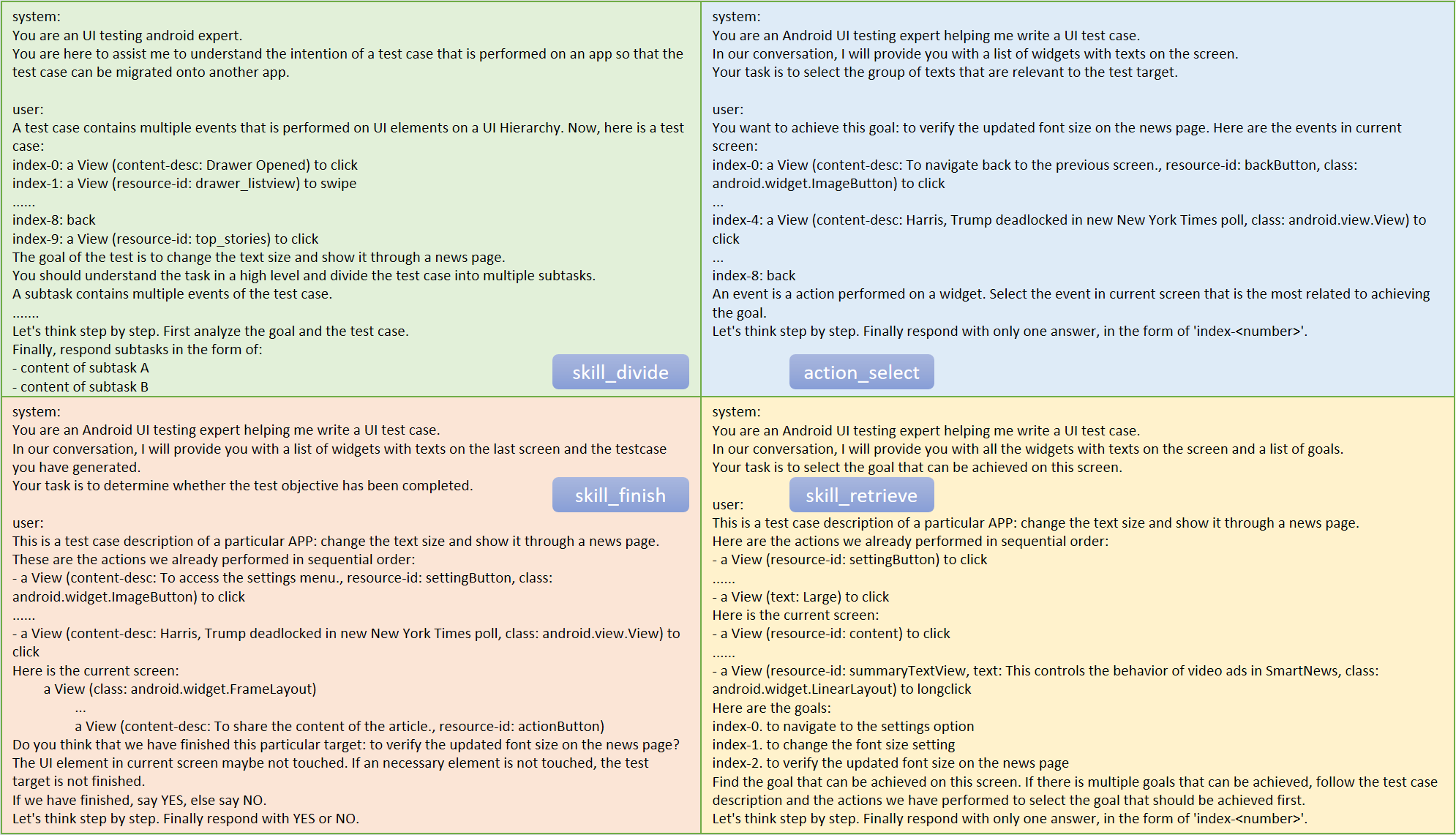}
    \caption{Prompt Design of \toolname{}.}
    \label{fig::prompt}
\end{figure*}
We translate the events and UI screens into natural language, then prompt the LLM to provide various types of responses. The prompt design is detailed in Figure~\ref{fig::prompt}.

\toolname{} provides five main UI actions for user interaction: navigating backward by pressing back, engaging with UI elements through clicking, and performing a long click by holding the center of an element for two seconds. Additionally, \toolname{} can execute swipe actions by querying the LLM for direction and input text by generating strings with the LLM for entry into UI elements. These features enhance navigation and manipulation of the user interface.

\section{Evaluation}\label{sec::eval}

\subsection{Research Questions}
To evaluate the effectiveness of \toolname{}, our evaluation focuses on answering the following research questions:
\begin{itemize}
    \item \textbf{RQ3}: How effective is \toolname{} against state-of-the-art UI test migration approaches? How general is \toolname{}?
    \item \textbf{RQ4}: How effective is the framework of \toolname{} for migrating UI test cases? How effective is the different module in \toolname{}?
\end{itemize}
We answer RQ3 by instancing \toolname{} with three LLMs and comparing them with state-of-the-art Android UI test migration tools. We also show the adaptiveness of \toolname{} in RQ3. 
We answer RQ4 by an ablation study.

\subsection{Evaluation Setup}\label{sec::setup}

\noindent \textbf{Dataset Construction.}
Several datasets, such as the ATM dataset~\cite{atm-dataset}, CraftDroid dataset~\cite{craftdroid-dataset}, and Fruiter dataset~\cite{fruiter-dataset} are available for evaluating UI test migration. However, these datasets have inconsistent formats. Additionally, due to the continuous upgrading of the Android platform, several test cases cannot be executed in our current environment. Therefore, we manually collected runnable test cases from apps that can be installed in our environment.

Our final dataset is a compilation of test cases from the aforementioned datasets~\cite{atm-dataset, craftdroid-dataset, fruiter-dataset}. It includes 103 test cases across 19 descriptions, resulting in 428 UI test migration pairs from source apps to target apps. The apps in our experimental subjects span seven categories, offering a broader range of app categories compared to those used in related studies~\cite{behrang2019test,hu2018appflow,lin2019test}.
The basic statistics of our experimental subjects are shown in Table~\ref{tab:benchmark-app}.

\begin{table}[t]
\begin{threeparttable}
\centering
\footnotesize
\caption{Overview of industrial apps used for evaluation}
\label{tab:benchmark-app}
\begin{tabular}{ccc|ccc}
\toprule
\textbf{Category} & \textbf{App Name}  & \textbf{\#Tasks} & \textbf{Category} & \textbf{App Name} & \textbf{\#Tasks}\\
\midrule
\multirow{5}{*}{\textbf{A2}} & a21  & 2 & \multirow{5}{*}{\textbf{A1}} & a11  & 2 \\
& a22  & 2 & & a12  & 2 \\
& a23  & 2 & & a13  & 2 \\
& a24  & 2 & & a14  & 2 \\
& a25  & 2 & & a15  & 2 \\
\cline{1-3}
\cline{4-6}
\multirow{5}{*}{\textbf{A5}} & a51  & 2 & \multirow{4}{*}{\textbf{Note}} & note2  & 1\\
& a52  & 2 & & note3  & 1\\
& a53  & 2 & & note4  & 1\\
\cline{4-6}
& a54  & 2 & \multirow{10}{*}{\textbf{News}} & ABC News  & 8\\
& a55  & 2 & & Buzzfeed$^*$  & 7\\
\cline{1-3}
\multirow{5}{*}{\textbf{Shop}} & shop1  & 1 & & Etsy  & 3\\
& shop2  & 1  & & Fox  & 10 \\
& shop3  & 1  & & Geek$^*$  & 5\\
& shop4  & 1  & & Home$^*$  & 5\\
\cline{1-3}
\multirow{4}{*}{\textbf{Expense}} & expense1  & 1 & & Reuters  & 3 \\
& expense2  & 1  & & SmartNews  & 7\\
& expense3  & 1  & & TheGuardian  & 11\\
& expense4  & 1  & & USAToday  & 6\\
\bottomrule
\end{tabular}
\begin{tablenotes}
\small
\item \textit{Apps with asterisks ($^*$) after their names are the ones that require logging in to access most features.}
\end{tablenotes}
\end{threeparttable}
\vspace{-10pt} 
\end{table}
\vspace{-0pt} 

\noindent \textbf{Test Platform.}
All experiments were conducted on official Android x64 emulators running Android 6.0, hosted on a server equipped with four AMD EPYC 7H12 64-Core Processors. Each emulator was allocated 4 dedicated CPU cores, 2 GB of RAM, and 2 GB of internal storage. To minimize mutual influences caused by disk I/O bottlenecks, emulator data were stored on an in-memory disk. Hardware graphics acceleration was enabled using two Nvidia GeForce RTX 3090 Graphics Cards to ensure the emulators' responsiveness. For apps requiring login to access evaluation features, we manually wrote auto-login scripts. Each script was executed only once before the corresponding app was tested in each test run.

\noindent\textbf{Evaluation Metrics.}
Unlike existing work, we adopt success rate as the evaluation metric due to the fundamental goal of migration is to mirror the functionality of source test case.
In UI test migration, given a source event from a source test case and a target app, previous formulation of migration approach attempts to find a matched target widget and synthesize a target event. The outcomes for a synthesized target event can be categorized into three types:
1. The synthesized target event matches an event in the ground truth (\emph{TP});
2. The synthesized target event does not match any event in the ground truth (\emph{FP}); and
3. An event in the ground truth is not found by the migration approach (\emph{FN}).
Previous studies~\cite{lin2019test, zhao2020fruiter, behrang2019test, zhang2024learning} primarily use three metrics—Precision, Recall, and F1-score—to measure the effectiveness of UI test migration.

However, we find that the Precision and Recall metrics do not accurately reflect the effectiveness of UI test migration tools. 
For instance, consider a task from the CraftDroid dataset: "Open web page `https://www.ics.uci.edu', open web page `https://uci.edu' and go back". If a tool generates the following test case:
\begin{itemize}
    \item Click the address bar
    \item Input `https://uci.edu' and press enter
    \item Click the address bar
    \item Input `https://www.ics.uci.edu' and press enter
    \item Go back
\end{itemize}

The Precision and Recall would both be 1. However, this test case does not correctly perform the intended events.
Moreover, we recognize that the primary goal of migration is to replicate the functionality of the source test case, rather than mapping the UI event sequence one by one.
Hence, we need a metric to verify the correctness of the generated test case.
To address this problem, we manually labeled an evaluator for each test case. The evaluator uses the UI screen sequence and event sequence to assess the functional correctness of the generated test case. We refer to this metric as Success Rate (SR). We use SR as the primary evaluation metric for the UI test migration task.

\subsection{RQ3: Overall Effectiveness of \toolname{}}

In this section, we assess the overall effectiveness of \toolname{}.

\noindent \textbf{Baseline Selection.}
For RQ3, we include CraftDroid as a baseline because it represents the state-of-the-art among publicly available tools with accessible source code and model parameters, enabling us to reproduce results within our dataset. As explained in Section \ref{subsec:prelim_rq2}, we do not include TEMDroid as a baseline. Given that SemFinder is specifically designed for event matching datasets, we integrate its best model into the CraftDroid framework by replacing the matching module, creating a hybrid baseline referred to as CraftDroid-SemFinder. Additionally, we consider the baselines from RQ2, namely CraftDroid-DeepSeek-V2 and CraftDroid-GPT-4o-mini.

For CraftDroid-based approaches, we set a time limit of 30 minutes, and found that approximately 20\% of the migrations reached this limit. We did not employ static analysis approaches, as they may be infeasible for some popular industry apps.

To evaluate \toolname{} against approaches utilizing LLMs, we instantiate \toolname{} with DeepSeek-V2 and GPT-4o-mini. Furthermore, to explore its capability in UI test migration, we instantiate \toolname{} with the advanced LLM GPT-4o.

\begin{table}[t]
\centering
\caption{Success rate of different migration approaches.}
\label{tab::rq3}
\renewcommand{\arraystretch}{1.2}
\begin{threeparttable}
\begin{tabular}{lccc}
\toprule
\textbf{Approach}  & \textbf{overall SR} & \textbf{1-to-1 mapping SR} & \textbf{non 1-to-1 mapping SR} \\
\midrule
\textbf{LLM Approaches} & & &\\
\cmidrule(r){1-1}
\toolname{}-DeepSeek-V2 & \textbf{63.6} & \textbf{80.1} & 53.0\\
\toolname{}-GPT-4o & 62.9 & 74.2 & \textbf{55.6}\\
\toolname{}-GPT-4o-mini & 58.4 & 75.5 & 47.4\\
CraftDroid-DeepSeek-V2 & 28.8 & 46.4 & 17.5\\
CraftDroid-GPT-4o-mini & 27.3 & 39.7 & 19.2\\
\midrule
\textbf{Non-LLM Approches} & & & \\
\cmidrule(r){1-1}
CraftDroid\cite{lin2019test}& 25.5 & 43.0 & 14.1\\
CraftDroid-SemFinder & 23.9 & 37.1 & 15.4\\
\bottomrule
\end{tabular}
\end{threeparttable}
\end{table}

Table~\ref{tab::rq3} shows the overall effectiveness of \toolname{} compared to other baseline approaches. \toolname{} demonstrates superior performance, particularly when paired with DeepSeek-V2, achieving the highest success rate (SR) of 63.6\%, closely followed by its combination with GPT-4o at 62.9\%. In contrast, CraftDroid has an SR of only 25.5\%.
When integrating SemFinder's best model into CraftDroid, we observe no increase in the success rate. This outcome highlights the gap between UI event matching, for which SemFinder is specifically designed, and UI test migration tasks.
Even when the mapping module is replaced with LLM, the improvement is minimal, reaching only 28.8\%.
Compared to the previous state-of-the-art approach CraftDroid, \toolname{} achieves a 149\% higher success rate. This highlights \toolname{}'s ability to effectively leverage LLMs to enhance UI test migration success, outperforming both its baseline and other state-of-the-art approaches.

To further analyze \toolname{}'s adaptability, we divided our dataset into two splits: a 1-to-1 mapping split, where events in the source and target test cases can be directly mapped, and a non 1-to-1 mapping split, which involves more complex scenarios as described in Section \ref{subsec:prelim_rq2}. For 1-to-1 mappings, \toolname{} achieves an SR improvement of 86\% over the state-of-the-art approach CraftDroid, highlighting its strength in scenarios with direct mappings. In non 1-to-1 mapping scenarios, CraftDroid's SR drops to 14.1\%. However, \toolname{} remains robust, with GPT-4o achieving a 55.6\% SR, representing a 294\% improvement over CraftDroid, demonstrating its ability to handle more complex migrations where direct mappings are not feasible.

The generalization capability of \toolname{} is evident as it consistently performs well across different LLM bases. Whether instantiated with DeepSeek-V2, GPT-4o, or GPT-4o-mini, \toolname{} maintains high success rates, underscoring its flexibility and effectiveness across various LLM implementations.

\noindent\fcolorbox{black}{gray!20}{
\begin{minipage}{\dimexpr\linewidth-2\fboxrule-2\fboxsep\relax}\Answer{RQ3}{
\toolname{} outperforms state-of-the-art approaches with a 149\% increase in overall SR. It addresses the challenge of non 1-to-1 mapping, achieving a 294\% improvement over state-of-the-art approaches in such scenarios. \toolname{} generalizes well with different LLMs.
}
\end{minipage}
}

\subsection{RQ4: Usefulness of the Imitation Learning Framework}

In this section, we assess the effectiveness of the \toolname{} framework through an ablation study.

\noindent \textbf{Baseline Selection.}
The first baseline, named TraceDroid, utilizes the entire source test case, the available event list in the target app, and the executed event history as inputs to the LLM. It prompts the model to select relevant events for UI test migration and uses the current UI screen to determine when migration should be completed. We use TraceDroid to evaluate the impact of incorporating skill and goal understanding in \toolname{}.
The second baseline, called TargetDroid, first comprehends the goal of the source test case. It uses this goal to guide test generation on the target app and decide when to finish, serving as an ablation to assess the usefulness of using skills as in-context examples.
CraftDroid with LLM replaces the event matching module of CraftDroid, serving as an ablation to determine the overall utility of the \toolname{} framework.

\begin{table}[t]
\centering
\caption{Ablation study of \toolname{}.}
\label{tab::rq4}
\renewcommand{\arraystretch}{1.2}
\begin{threeparttable}
\begin{tabular}{lccc}
\toprule
\textbf{Approach}  & \textbf{overall SR} & \textbf{1-to-1 mapping SR} & \textbf{non 1-to-1 mapping SR} \\
\midrule
\toolname{}-DeepSeek-V2 & \textbf{63.6} & \textbf{80.1} & \textbf{53.0}\\
TraceDroid-DeepSeek-V2 & 32.5 & 41.1 & 26.9 \\
TargetDroid-DeepSeek-V2 & 48.6 & 62.9 & 39.3 \\
CraftDroid-DeepSeek-V2 & 28.8 & 46.4 & 17.5\\
\midrule
\toolname{}-GPT-4o-mini & \textbf{58.4} & \textbf{75.5} & \textbf{47.4}\\
TraceDroid-GPT-4o-mini & 27.5 & 43.0 & 17.5 \\
TargetDroid-GPT-4o-mini & 45.2 & 62.3 & 34.2 \\
CraftDroid-GPT-4o-mini & 27.3 & 39.7 & 19.2\\
\bottomrule
\end{tabular}
\end{threeparttable}
\end{table}

The ablation study in Table~\ref{tab::rq4} highlights the effectiveness of the \toolname{} framework compared to several baseline methods. \toolname{} significantly surpasses TraceDroid, TargetDroid, and CraftDroid in SR, underscoring the importance of integrating skill and goal understanding to improve UI test migration.
Notably, TargetDroid outperforms both CraftDroid and TraceDroid, emphasizing the critical role of goal comprehension in successful test generation. As explained in Section~\ref{sec::problem}, the objective of UI test migration is to replicate the functionality on the AUT with an equivalent event sequence, rather than transferring events one by one. TargetDroid's emphasis on maintaining a stable goal effectively guides the test generation process across different UI states.

\noindent\fcolorbox{black}{gray!20}{
\begin{minipage}{\dimexpr\linewidth-2\fboxrule-2\fboxsep\relax}\Answer{RQ4}{
The \toolname{} framework is highly effective for migrating UI test cases, outperforming TraceDroid, TargetDroid, and CraftDroid in the ablation study. Its success is attributed to the integration of skill and goal understanding, which enhances adaptability and accuracy in complex scenarios.
}
\end{minipage}
}
\section{Threats to Validity}
The main external threats to the validity include the extent to which the subject apps and tools selected for our evaluation are representative of true practice.
To mitigate the impact of the bias introduced by app selection, we use highly popular industrial apps widely used by related work.
To mitigate the impact of the bias introduced by tool selection, we choose the latest and widely used state-of-the-art and state-of-the-practice automated UI testing tools for comparison.

The threats to internal validity are instrumentation effects that can bias our results, including faults in our implementation of \toolname{}, parameter selection of \toolname{}.
To reduce these threats, We carefully test and validate \toolname{} to assure the behavior of \toolname{}.
\section{Related Work}
\noindent\textbf{Automated UI testing.} Automated UI testing has long been popular within the research community~\cite{AndroidMonkey,mao2016sapienz,su2017guided,Choi:2013,Hao:2014,yang13:grey,Machiry:2013,Ye:2013:DFA:2536853.2536881,DroidBot,Azim:2013,EvoDroid,Anand:2012,Gu:2019:PGT:3339505.3339542,dong2020timemachine,pan2020qtesting,romdhana2022deep} and industry~\cite{mao2016sapienz,zheng2017automated,ran2022automated}. 
Current automated UI testing tools can largely be categorized into four groups:
(1) Some tools generate random test inputs and/or use evolutionary algorithms to evolve these test inputs~\cite{AndroidMonkey,mao2016sapienz,Machiry:2013,Ye:2013:DFA:2536853.2536881,dong2020timemachine}.  
For example, Monkey~\cite{AndroidMonkey} generate pseudo-random actions such as clicks and drags to brute-force testing the app. 
(2) Some tools explore the app under test systematically~\cite{Azim:2013,EvoDroid,Anand:2012}.
(3) Model-based tools construct a \textit{UI transition model} to track history exploration, decide current progress, and plan for actions triggering the parts of the app that have not been explored~\cite{Choi:2013,Hao:2014,su2017guided,DroidBot,Gu:2019:PGT:3339505.3339542}.
(4) Some tools also depends on machine learning approachs such as deep learning or reinforcement learning techniques to guide the exploration of the app~\cite{koroglu2018qbe,li2019humanoid,pan2020qtesting,romdhana2022deep}.
As an example, Q-testing~\cite{pan2020qtesting} explores toward rare UI screen using reinforcement learning based on curiosity.
Comparing to test migration, automated UI testing research has largely devoted to achieving high coverage and triggering certain bugs hidden in the app, while test migration aims to generate test cases explicitly from an existing test case on another app.

\noindent \textbf{Android Test Migration.}
The main focus of existing work is to find a gold mapping for widget-level migration.  
Appflow~\cite{hu2018appflow} trains a classifier model to distinguish the type of the widgets (e.g., a widget that leads to settings) on the current screen and in the source test.
A pair of widgets in the source test and on the UI screen of target app with the same type is deemed as similar and suitable for migration.
There are also several approaches~\cite{behrang2019test,lin2019test,mariani2021evolutionary,zhang2024learning} that view test migration as a form of matching task, where the goal is to find widgets on the target app that match one of the widgets in the source test.
ATM~\cite{behrang2019test} first extracts all the textual information for the widgets on the current screen that is the one waiting for matching in the source app.
Then, ATM utilizes a fine-tuned version of word2vec~\cite{mikolov2013distributed} to calculate the similarity between each pair of texts in the gathered textual information, after which the similarity score is aggregated by average into the similarity between widgets.
Based on the similarity score, ATM selects the pair with the highest similarity that is above a given threshold as the candidate for migration.
If no such pair of widgets can be found on the current screen, it tries to search on other explored GUI states of the app for such pairs until a certain time limit runs out.
Craftdroid~\cite{lin2019test} adopts a similar approach, but uses both the textual similarity and the distance between the candidate target widget for migration and the previous migrated widget on the GUI state transition graph to measure the similarity.
Adaptdroid~\cite{mariani2021evolutionary}, also similar, uses a word-move model~\cite{kusner2015word} (another word embedding model) to calculate the similarity of two widgets.
TEMDroid~\cite{zhang2024learning} uses a more modern and robust language model BERT\cite{devlin2018bert} and fine-tunes it to achieve a more accurate similarity evaluation.
In comparison, our work does not fall in line and focuses on non one-to-one mapping and designs better workflow of UI test migration, which is orthogonal to existing work.

\noindent \textbf{Imitation Learning.} Imitation learning often involves an agent learning to mimic expert behavior after given demonstrations from the expert~\cite{zheng2021imitation,zare2024survey,hussein2017imitation,ho2016generative,duan2017one,osa2018algorithmic}.
Imitation learning (IL), when applied to UI test migration, can be used to learn from source test the expertise from human developer and transfer that knowledge onto new apps.
There are two main approaches for IL: Behavior Cloning (BC)~\cite{torabi2018behavioral,florence2022implicit,codevilla2019exploring,ly2020learning,zhan2018generative,bratko1995behavioural}, which attempt to map the actions in the demonstration directly into a new environment, and Inverse Reinforcement Learning (IRL)~\cite{arora2021survey,ng2000algorithms,zhifei2012survey,hadfield2016cooperative,ramachandran2007bayesian,ziebart2008maximum}, where the agent tries to deduce the hidden value function ~\cite{eschmann2021reward,icarte2022reward,matignon2006reward,jin2020reward} that leads to the expert's behavior. 
This method is helpful for adapting demonstrations to different contexts.
Under the context of UI testing, the IRL method helps migrating each test actions to different UI contexts, which facilitates the transfer of test cases to other apps.
\toolname{} uses IL to migrate test cases by imitating what is carried out in the source test case, which is the expert demonstration in this case, and learns to transfer the knowledge to a new app.
Different from the widget-level matching technique that traditional methods like ATM~\cite{behrang2019test} or Craftdroid~\cite{lin2019test} adopt, \toolname{} leverages IL alongside large language models (LLM) to generate and adapt action sequences based on contextual understanding~\cite{ghosh2016contextual,smith2019contextual}. 
This makes \toolname{} more flexible and effective in handling varied UI layouts and behaviors across different apps.

\noindent\textbf{Large Language Models and Their Applications.}
Large language models~\cite{brown2020language,codex} is a type of artificial intelligence (AI) that has gained significant attention and popularity in recent years. 
These models are designed and trained to deal with natural language tasks, enabling them to excel at a wide range of natural language process (NLP) tasks like language translation~\cite{stahlberg2020neural}, question-answering~\cite{namazifar2021language}, and text summarization~\cite{brown2020language}. 
OpenAI's GPT-3~\cite{brown2020language} is one of the many most well-known instances of large language models that consists of 175 billion parameters.
This also makes it one of the largest models up to date.
Large language models have been also widely used for other software engineering tasks such as UI understanding~\cite{wang2023enabling}, code generation~\cite{codex}, string input generation~\cite{liu2022fill}, defect prediction~\cite{pan2021empirical}, and unit test generation~\cite{schafer2023adaptive}.

\section{Conclusion}
UI test migration is promising to automatically generate high-quality test cases.
Existing UI test migration approaches heavily rely on advanced machine learning models to map UI events between apps, often overlooking the functionality of UI test migration. As a result, these approaches are limited to scenarios with direct 1-to-1 sequential mapping and are ineffective for real-world apps.
To address the preceding issues, in this paper, we have formulated the UI test migration into a skill-adaptive imitation learning problem, which focuses on mirroring the functionality of source test cases. 
Based on our formulation, we have proposed the \toolname{} framework and instantiated it into a train-free practical approach for Android UI test migration.
\toolname{} complements the existing literature and emphasizes the importance of retaining high-level functional skills instead of solely improving event mapping accuracy.
Extensive evaluations on \appNum{} mobile apps have demonstrated the effectiveness of \toolname{}. Compared to state-of-the-art Android UI test migration approaches, \toolname{} achieves a 149\% improvement in overall success rate and an impressive 294\% increase in cases where the source and target test cases lack direct 1-to-1 mapping.

\balance
\bibliographystyle{ACM-Reference-Format}
\bibliography{ref}

\end{document}